\documentclass[sn-mathphys,Numbered]{sn-jnl}

\usepackage{graphicx}%
\usepackage{multirow}%
\usepackage{amsmath,amssymb,amsfonts}%
\usepackage{amsthm}%
\usepackage{mathrsfs}%
\usepackage[title]{appendix}%
\usepackage{xcolor}%
\usepackage{textcomp}%
\usepackage{manyfoot}%
\usepackage{booktabs}%
\usepackage{subcaption}%
\usepackage[font={small}]{caption}
\usepackage{algorithm}%
\usepackage{algorithmicx}%
\usepackage{algpseudocode}%
\usepackage{listings}%
\usepackage{braket}%
\usepackage{subcaption}%
\usepackage{qcircuit}%
\usepackage{bm}%
\usepackage{bbm}
\usepackage{array}
\usepackage{breqn}%
\usepackage[export]{adjustbox}
\newcommand{\etal}{\textit{et al. }}
\usepackage[section]{placeins}
\DeclareMathOperator*{\Motimes}{\text{\raisebox{0.25ex}{\scalebox{0.8}{$\bigotimes$}}}}

\begin{document}

\title[Article Title]{Quantum Convolutional Neural Networks for Multi-Channel Supervised Learning}


\author[1]{\fnm{Anthony M.} \sur{Smaldone}}\email{anthony.smaldone@yale.edu}

\author[1]{\fnm{Gregory W.} \sur{Kyro}}\email{gregory.kyro@yale.edu}

\author*[1]{\fnm{Victor S.} \sur{Batista}}\email{victor.batista@yale.edu}

\affil[1]{\orgdiv{Department of Chemistry}, \orgname{Yale University}, \orgaddress{\street{225 Prospect Street}, \city{New Haven}, \postcode{06511}, \state{CT}, \country{USA}}}


\abstract{As the rapidly evolving field of machine learning continues to produce incredibly useful tools and models, the potential for quantum computing to provide speed up for machine learning algorithms is becoming increasingly desirable. In particular, quantum circuits in place of classical convolutional filters for image detection-based tasks are being investigated for the ability to exploit quantum advantage. However, these attempts, referred to as quantum convolutional neural networks (QCNNs), lack the ability to efficiently process data with multiple channels and therefore are limited to relatively simple inputs. In this work, we present a variety of hardware-adaptable quantum circuit ansatzes for use as convolutional kernels, and demonstrate that the quantum neural networks we report outperform existing QCNNs on classification tasks involving multi-channel data. We envision that the ability of these implementations to effectively learn inter-channel information will allow quantum machine learning methods to operate with more complex data.}

\keywords{Quantum Machine Learning, Convolutional Neural Networks, Multi-Channel Data, Image Classification, Supervised Learning}



\maketitle

\section{Introduction}\label{sec1}

Quantum computers are devices that utilize quantum mechanical phenomena such as superposition and entanglement to solve problems that are infeasible with timescales provided by classical computers. From the time quantum computers were first pro- posed in 1982 \cite{Feynman1982}, researchers have been working to develop algorithms that provide quantum advantage over classical algorithms \cite{Grover1996,Shor1994,Deutsch1992,Cleve1998,Bernstein1997,Simon1997,Kitaev1995}. The main application challenges are related to noise, high gate errors and short decoherence times \cite{Preskill2018}, necessitating the development of more efficient quantum circuit ansatzes. In recent years, devices have demonstrated quantum supremacy such as those introduced by Google AI Quantum (2019) \cite{Arute2019}, the University of Science and Technology of China (2020) \cite{Zhong2020}, and IBM (2021) \cite{chow2021ibm}.

Machine learning (ML) has been gaining significant attention in recent years for strongly influencing many important fields of study and sectors of society. Image recognition \cite{Pak2017,Wu2015,Liu2020}, natural language processing  \cite{otter2020survey,young2018recent,li2018deep}, self-driving vehicles \cite{FUJIYOSHI2019244,gupta2021deep,rao2018deep,daily2017self,mozaffari2020deep}, robotics \cite{Pierson2017,wang2019artificial,kim2021review,lesort2020continual,kleeberger2020survey}, and molecular biochemistry \cite{Choi2019,Bonetta2019,Ballester2010,Sino_2021} are just some of the fields that are being revolutionized by ML. The most prominent example would be ChatGPT, a chatbot built on top of OpenAI’s generative pre-trained transformer (GPT)-3.5 and GPT-4 large language models that have been fine-tuned for conversation using both supervised and reinforcement learning from human feedback techniques \cite{gpt35,gpt4}. In the natural sciences, Deepmind’s Alphafold is able to accurately predict 3D models of protein structures, and is accelerating research in nearly every field of molecular biology and biochemistry \cite{Jumper2021}.

The advances in both quantum computing and ML inspire the development of quantum ML (QML) methods to exploit the speed of quantum computations and the predictive capabilities of ML. There has been recent work demonstrating the feasibility and advantages of substituting components of classical ML architectures with quantum analogues \cite{schuld2015introduction,biamonte2017quantum,zhang2020recent,Cerezo2022}, such as quantum circuits in place of classical convolutional kernels in convolutional neural networks (CNNs) \cite{Cong2019,Henderson2019,Oh2020,Chen2022,Hong2021,Jing2022,Mishra2023,Mari_2021,Hur2022}. Classical CNNs are state-of-the-art for image, video, and sound recognition tasks \cite{Rawat2017,LeCun2015} and also have applications in the natural sciences \cite{Jumper2021, Gao2021, Wei2019,Chen2018,Kyro2023,Casey2020,Senior2020}. CNNs that incorporate quantum circuits to function as kernels, referred to as Quantum CNNs (QCNNs), have performed well on classification tasks involving simple data such as the MNIST dataset of handwritten digits \cite{Hur2022,Oh2020}, as well as multi-channel data such as the CIFAR-10 dataset \cite{Jing2022,Riaz2023}.

The QCNN was introduced in 2019 by Cong \etal  \cite{Cong2019}, and was applied to quantum phase recognition and quantum error correction optimization. Since then, QCNNs have been applied in areas ranging from high energy physics \cite{Chen2022} to biochemistry \cite{Hong2021}. However, current methods either do not effectively capture inter-channel information, or require more qubits than are currently permissible, and therefore lack the ability to efficiently process more complex data with multiple channels.

With current QCNNs, the number of required qubits scales linearly with the length of the channel dimension of the input data. This is a feasible approach for simple data that can be modeled with small filters. For instance, Jing \etal  demonstrated the ability to use a 12- and 18-qubit circuit to function as a 2 $\times$ 2 and 3 $\times$ 3 convolutional filter, respectively, on low resolution red-green-blue (RGB) images (three channels, one for each color) \cite{Jing2022}. However, extension of this approach to tasks involving data with more channels is prohibited by current hardware limitations. As an attempt to overcome this challenge, there has been much recent work that performs a measurement on each channel individually, collapsing the wavefunction after measuring a given channel of the data and storing the measurement classically \cite{Chen2022,Hong2021,Oh2020,Mishra2023}. Although the hardware requirements have no dependence on the number of channels when using this method, much of the inter-channel information is lost, which is valuable for accurately modeling the data. In this work, we propose several methods for operating with multi-channel data that preserve inter-channel information and require a number of qubits that is independent of the length of the channel dimension of the input data.

\section{Quantum Computers for Convolutional Neural Networks}\label{sec2}
\subsection{Basics of Quantum Computing}
Qubits are two-level systems used in computations to exploit the quantum mechanical phenomena of superposition and entanglement. The state of a single qubit can be represented by:
\begin{equation}
   \ket{\psi} = \alpha\ket{0}+\beta\ket{1} \text{ such that } \lvert\alpha\rvert^2 + \lvert\beta\rvert^2 = 1 \text{ and } \alpha, \beta \in \mathbbm{C}.
\end{equation}

The state of an $n$-qubit system is represented by the superposition of all possible n-bit strings:

\begin{equation}
    \ket{\psi} = \sum_{x \in \{0,1\}^n}\alpha_x\ket{x}
\end{equation}

These qubit states are transformed via unitary operations. For any single qubit unitary,
\begin{equation}
    U = \begin{bmatrix}
        u_{00} & u_{01} \\
        u_{10} & u_{11} \\
    \end{bmatrix},
\end{equation}

the matrix of a two-qubit controlled operation may be expressed as 
\begin{equation}
CU = 
    \begin{bmatrix}
        1 & 0 & 0 & 0 \\
        0 & 1 & 0 & 0 \\
        0 & 0 & u_{00} & u_{01} \\
        0 & 0 & u_{10} & u_{11} \\ 
    \end{bmatrix}.
\end{equation}

In a two-qubit controlled gate, if the control qubit is in the $\ket{1}$ state, the unitary operation is applied to the target qubit. If the control qubit is in the $\ket{0}$ state, the target qubit is unaffected. Measurements on a quantum state are performed by taking the expectation value of a Pauli operator, $\sigma$ as shown in Equation \ref{measurement}. Quantum circuits model the sequential manipulation and measurement of qubits.

\begin{equation}\label{measurement}
    \mathcal{M} = \bra{\psi}\sigma_{x,y,z}\ket{\psi}
\end{equation}

\subsection{The Structure of Quantum Convolutional Neural Networks}

\begin{figure}
    \centering
\begin{tabular*}{\linewidth}{*{2}{>{\centering\arraybackslash}p{\dimexpr0.5\linewidth-2\tabcolsep}}}
\includegraphics[width=\linewidth,valign=c]{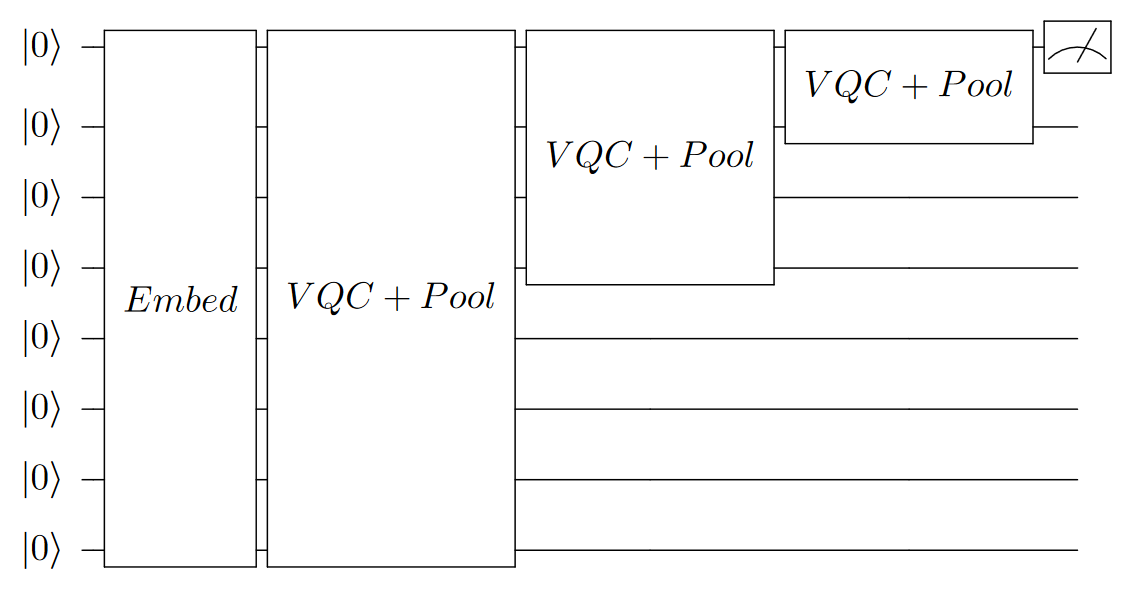}
    &   \includegraphics[width=1.0\linewidth,valign=c]{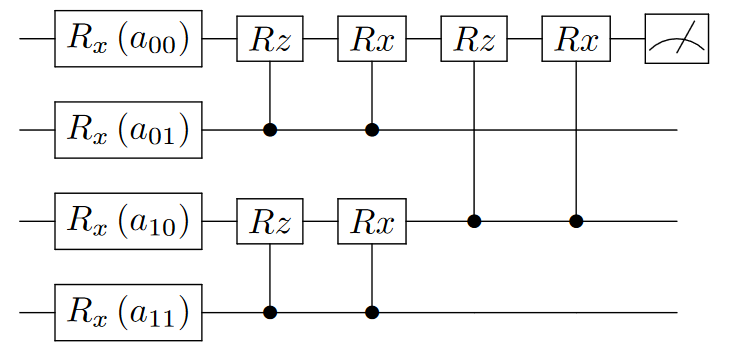}   \\
\caption{A general circuit for a quantum convolutional neural network with 8 input qubits. The classical data is embedded, followed by variational quantum circuits and pooling layers, concluded with a measurement.}  \label{trad_QCNN}
    &   \caption{The simple circuit for a quantum convolutional neural network demonstrated by Oh \etal \cite{Oh2020} which inspired the Control model \cite{Menborong2020}.}  \label{Oh2020_QCNN}
    \end{tabular*}
\end{figure}

The traditional structure of QCNNs consist of embedding classical data in a quantum mechanical state, having $O\left(\log\left(n\right)\right)$ convolutional and pooling layers for $n$ initial qubits, a measurement layer, followed by a classical fully-connected layer. The ``convolutional" component in this quantum circuit is a variational quantum circuit (VQC). A VQC is a series of gates such that they have learnable parameters that are updated via backpropagation. The circuit architecture for a traditional QCNN is shown in Figure \ref{trad_QCNN}. While not a hard requirement, this hierarchical structure of halving the number of qubits with each layer (analogous to classical pooling in CNNs) circumvents the ``barren plateau" phenomenon and thus guarantees trainability. Oh \etal \cite{Oh2020} provides a simple QCNN tutorial for a 2 $\times$ 2 filter, utilizing a single VQC and pooling layer shown in Figure \ref{Oh2020_QCNN}. This structure was implemented \cite{Menborong2020} and is used as the comparative ``Control" model with respect to this work.

\subsection{Multi-Channel Data with Quantum Convolutional Neural Networks}
\begin{figure}[H]
  \centering
  \includegraphics[width=0.35\textwidth]{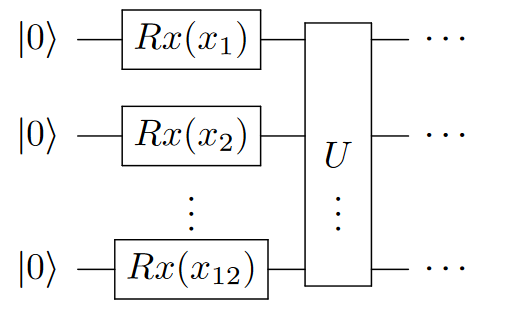}
  \caption{{The embedding and convolutional scheme for Jing \etal's \cite{Jing2022} implementation of a 2 $\times$ 2 filter on a color image. The twelve pixels covered by the filter are flattened and angle encoded via a rotation about the Pauli X axis, and then passed to a variational unitary block $U$.}}
  \label{fig:Jing2022}
\end{figure}

Previous works \cite{Oh2020,Chen2022,Hong2021,Mishra2023} have performed VQCs on each channel, while storing this information classically between channel convolutions and summing the result. Collapsing the wavefunction with a measurement between channels breaks any entanglement between them, hindering the model's ability to learn inter-channel patterns effectively. To our knowledge, the only attempt at including all channel data into a quantum circuit for convolutions was performed by Jing \etal \cite{Jing2022}. Jing takes an intuitive approach and increases the number of qubits proportionally to the number of pixels covered by the filter. Thus, a 2 $\times$ 2 filter acting on an image containing three color channels requires 12 qubits in this approach, as shown in Figure \ref{fig:Jing2022}.

\section{Extending QCNNs for Multi-Channel Data}

\subsection{The Channel Overwrite Method}
The proposed Channel Overwrite (CO) method takes advantage of a controlled phase gate applied to a single ancilla qubit to entangle information between channels. In our case, a single channel of classical data is encoded into a quantum state and passed through a learnable set of unitary operations. Quantum information from the working qubits is exchanged onto the ancilla qubit via a controlled phase gate, where the phase shift angle is a learnable parameter. The classical data for the next channel is angle encoded onto the working qubits, passed though the learnable set of unitaries, and has information further exchanged with the ancilla using a controlled phase gate. This process is repeated for each channel. After processing all channels of the data, the ancilla qubit is measured, and the expectation value is used as the output of that quantum convolution.

\subsubsection{State Preparation}\label{subsubsec2}
The CO-QCNN method removes the dependency of the number of necessary qubits on the length of the channel dimension of the data, and requires only $F^2 + 1$ qubits, where $F$ is the length of a single dimension of the square filter. The working qubits perform a convolution by executing a set number of learnable unitary gates over one channel of an input signal. The resulting state undergoes a phase shift controlled by the ancilla qubit.
The uninitialized quantum state and classical input can be represented as:

\begin{equation}\label{uninit}
    \ket{\Psi_0} = \ket{0} \otimes \ket{\psi_0} = \ket{0} \otimes \ket{0}^{\otimes F^2}, \bold{x}_{l,w,c} \in \mathbb{R}^{N-1}
\end{equation}

where $l, w, c$ are the initial indices of an input 3D-tensor of total length $L$, width $W$, and channels $C$, respectively, $\ket{\Psi_0}$ is the state of the total quantum system, and $\ket{\psi_0}$ is the state of the working qubits. $\bold{x}_{l,w,c}$ is a vector of the flattened normalized input data  covered by the filter for a single channel, described by the expression:
\begin{equation}
    \bold{x}_{l,w,c} = \{x_{l,w,c},x_{l,w+1,c},\cdots,x_{l,w',c}, \\x_{l+1,w,c},x_{l+1,w+1,c},\cdots,x_{l+1,w',c},\cdots,x_{l',w',c}\}
\end{equation}

where $l'$ and $w'$ represent the final index covered by the filter of the length and width dimensions.

\begin{equation}
    l' = l+F-1, w'=w+F-1
\end{equation}

In this work, the input data is encoded into a quantum state with angle encoding, as this method requires only a single gate per qubit and permits a low circuit depth compared to other encoding schemes. We encode all normalized data $\bold{x}$ with a rotation about the Pauli-X axis by $\pi x_i$ radians. Additionally, we prepare the ancilla qubit by placing it into a state of maximal superposition with a Hadamard transformation. The matrix representation of the gates for a rotation about the Pauli-X axis and a Hadamard transformation are described in Equations \ref{Rx} and \ref{Hadamard}, respectively.

\begin{equation}\label{Rx}
\bold{R_x}\left(\theta\right) =
    \begin{bmatrix}
        \cos\left(\frac{\theta}{2}\right) & -\sin\left(\frac{\theta}{2}\right) \\
        -\sin\left(\frac{\theta}{2}\right) & \cos\left(\frac{\theta}{2}\right) \\
    \end{bmatrix}
\end{equation}
\begin{equation}\label{Hadamard}
\bold{H} = 
\frac{1}{\sqrt{2}}
    \begin{bmatrix}
        1 & 1 \\
        1 & -1 \\
    \end{bmatrix}
\end{equation}

Letting $U_{RX,c}$ be the unitary block that embeds the classical data of channel $c$, the prepared quantum state is represented as:
\begin{equation}\label{angle_encode}
    \ket{\Psi_1} = \ket{+} \otimes \ket{\psi_1} = \bold{H}\ket{0} \otimes \Motimes_{i=1}^{F^2} \bold{R_x}\left(x_i \cdot \pi \right)\ket{\psi_0} = \bold{H}\ket{0} \otimes U_{RX, 1} \ket{\psi_0}
\end{equation}

\subsubsection{Circuit Construction}
Let $\bold{U}$ be the set of $\gamma$ chosen unitaries to perform the quantum convolution.
\begin{equation}\label{eq:8}
    \bold{U} = \prod_{i=1}^\gamma U_i
\end{equation}

After the classical data has been encoded and the deposition qubit has been placed into a state of maximal superposition, $\bold{U}$ is applied to the working qubits, transforming the total state according to:

\begin{equation}\label{eq:14}
    \ket{\Psi_2} = \ket{+} \otimes \bold{U} \ket{\psi_1} = \ket{+} \otimes \ket{\psi_2}.
\end{equation}

Following this learnable unitary block, we parametrically exchange phasic information from the entangled working register and the prepared ancilla qubit. This is achieved via a controlled phase gate, $CP$:

\begin{equation}\label{eq:15}
    \ket{\Psi_3} = CP\left(\ket{+} \otimes \ket{\psi_2}\right),
\end{equation}

where the matrix representation of the single qubit phase gate is:

\begin{equation}
P\left(\theta\right) =
    \begin{bmatrix}
        1 & 0 \\
        0 & e^{i\theta}
    \end{bmatrix}  
\end{equation}

It should be noted that this exchange of phasic information with the ancilla qubit is ideally applied to each qubit in the working register to retain as much information as possible. However, connecting all qubits to a single particular qubit can be challenging depending on the available quantum hardware. Unless otherwise stated, all results shown in this work restrict the controlled operations to the ancilla qubit and the first qubit of a given working register.

\begin{figure}
  \centering
  \includegraphics[width=1.0\textwidth]{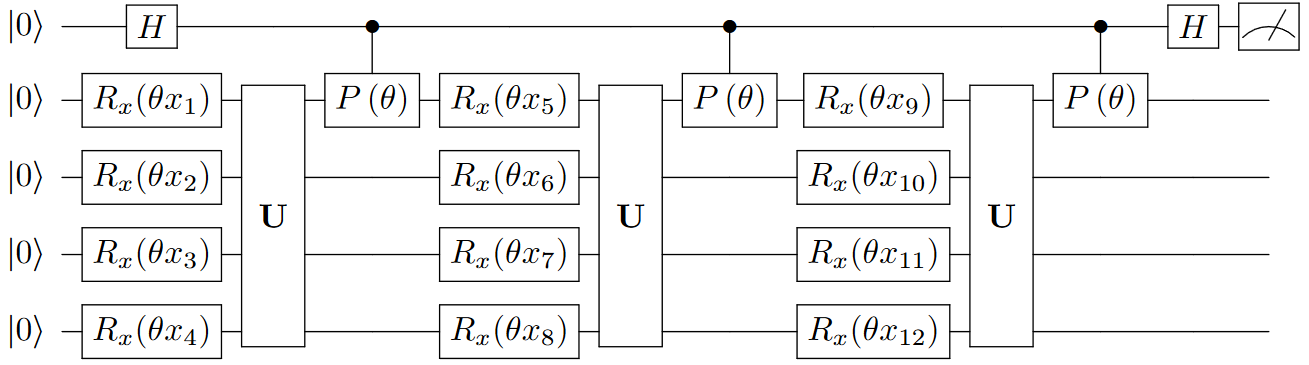}
  \caption{Channel Overwrite (CO) circuit using a 2 $\times$ 2 filter applied to three channels.}
  \label{fig:DR_circuit}
\end{figure}

The working qubits are then overwritten with data from the next channel, and the circuit proceeds as before. This procedure of angle encoding channel data and executing a controlled phase gate with the ancilla is repeated for each channel of the input data. Therefore the final state $\ket{\Psi_f}$ of the circuit is described by Equation \ref{full_circuit}, and output of this circuit is obtain via a measurement (Equation \ref{measure}) of the ancilla qubit in the Hadamard basis. A sample circuit is shown in Figure \ref{fig:DR_circuit}.

\begin{equation}\label{full_circuit}
    \ket{\Psi_f} = \left(\prod_{c=2}^{C}CP \cdot \bold{U} \cdot U_{RX,c}\right)\ket{\Psi_3} 
\end{equation}

\begin{equation}\label{measure}
    \mathcal{M} = \bra{\bold{H}\Psi_f} \sigma_z \ket{\bold{H}\Psi_f}
\end{equation}

\subsection{Parallel Channel Overwrite Method}
The Parallel Channel Overwrite-QCNN (PCO-QCNN) method decreases the circuit depth compared to that which is used in the CO-QCNN method. The PCO-QCNN operates simultaneously on $R$ parallelized channels, where $R$ is the number of specified working registers. Information from these working registers is exchanged with the ancilla and are all overwritten with the next set of channel data as in the CO-QCNN method. This method therefore requires $RF^2+1$ qubits.

\subsubsection{State Preparation}
The states of the working qubits in the PCO-QCNN method are prepared similarly to the CO-QCNN method, with the caveat of each register being encoded with a different channel of the data The overall quantum state is initialized as done in Equation \ref{POC_QCNN_starting}, where $\ket{\psi_{0,r}}$ is the initial substate of qubits in a given register. The data covered by the filter of $R$ channels is angle encoded as done in Equation \ref{angle_encode_parallel}, and the ancilla qubit is placed in a uniform superposition.

\subsubsection{Circuit Construction}\label{DRPC_construction}

The method of applying a controlled phase gate and overwriting is extended to $R$ working qubit registers in the PCO-QCNN. First intra-channel data is entangled followed by the entanglement of inter-channel data. This hierarchical method of first entangling intra-channel data (Equation \ref{DRPC_intra}) and then inter-channel data (Equation \ref{DRPC_inter}) is similar to the method used in the HQConv circuit \cite{Jing2022}. Controlled phase gates are performed between the ancilla qubit and the first qubit of each working register in Equation \ref{DRPC_phase}, and a measurement is taken as previously done in Equation \ref{measure}. The state evolution of the PCO-QCNN circuit is sequentially presented in Equations \ref{POC_QCNN_starting}-\ref{PCO_final_state}.
\begin{equation}\label{POC_QCNN_starting}
    \ket{\Psi_0} = \ket{0} \otimes \Motimes_{r=1}^R \ket{\psi_{0,r}} = \ket{0} \otimes \ket{0}^{\otimes RF^2}
\end{equation}

\begin{equation}\label{angle_encode_parallel}
    \ket{\Psi_1} = \ket{+} \otimes \Motimes_{r=1}^R \ket{\psi_{1,r}} = \bold{H}\ket{0} \otimes \Motimes_{r=1}^{R} \left( U_{RX,r}\left(x_r \cdot \pi \right)\ket{\psi_{0,r}}\right)
\end{equation}

\begin{equation}\label{DRPC_intra}
    \ket{\Psi_2} = \ket{+} \otimes \Motimes_{r=1}^{R}\bold{U}\ket{\psi_{1,r}} = \ket{+} \otimes \Motimes_{r=1}^{R}\ket{\psi_{2,r}}
\end{equation}

\begin{equation}\label{DRPC_inter}
    \ket{\Psi_3} = \ket{+} \otimes \bold{U}_c \Motimes_{r=1}^{R}\ket{\psi_{2,r}} = \ket{+} \otimes \ket{\psi_3}
\end{equation}

\begin{equation}\label{DRPC_phase}
    \ket{\Psi_4} = CP^R \ket{\Psi_3}
\end{equation}

\begin{equation}\label{PCO_final_state}
    \ket{\Psi_f} = \left(\prod_{l=1}^{C/R}CP^R \cdot \bold{U}_c \cdot \Motimes_{r=l}^{l+R}\left(\bold{U} \cdot U_{RX,r}\right)\right)\ket{\Psi_4}
\end{equation}

\begin{figure}
  \centering
  \includegraphics[width=0.9\textwidth]{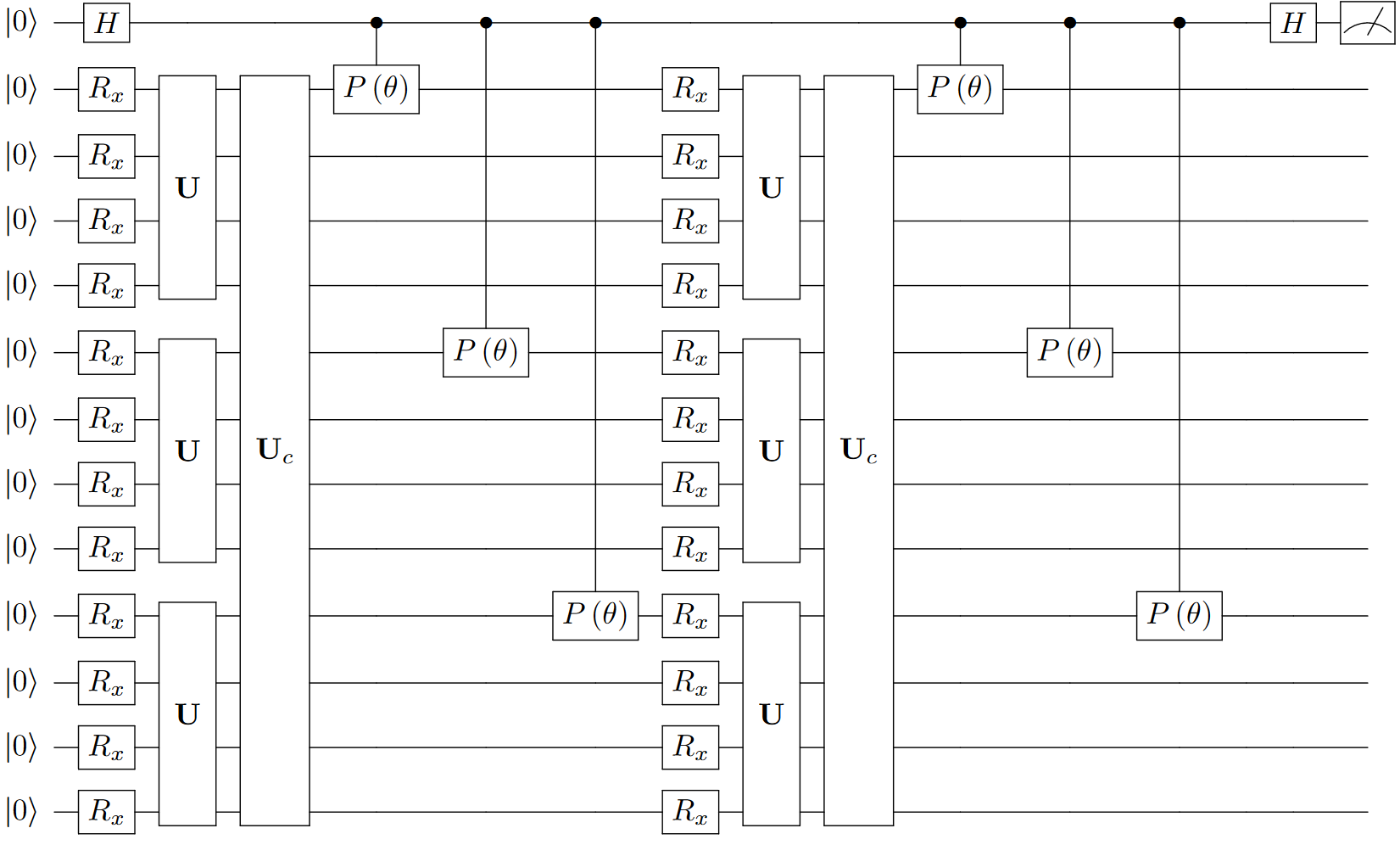}
  \caption{Parallel Channel Overwrite (PCO) circuit with three registers using a 2 $\times$ 2 filter applied to six channels. $\bold{U}_c$ (Appendix Figure \ref{fig:inter_channel_u}) is a block of unitary gates to entangle the different working qubit channel registers.}
  \label{fig:DRPC_circuit}
\end{figure}

This circuit is visualized in Figure \ref{fig:DRPC_circuit}. Although more qubits are being used than in the CO-QCNN, the number of gates and circuit depth have been reduced.

\subsection{Parallel Channel Overwrite - Topologically Considerate Method}
\begin{figure}
  \centering
  \includegraphics[width=0.9\textwidth]{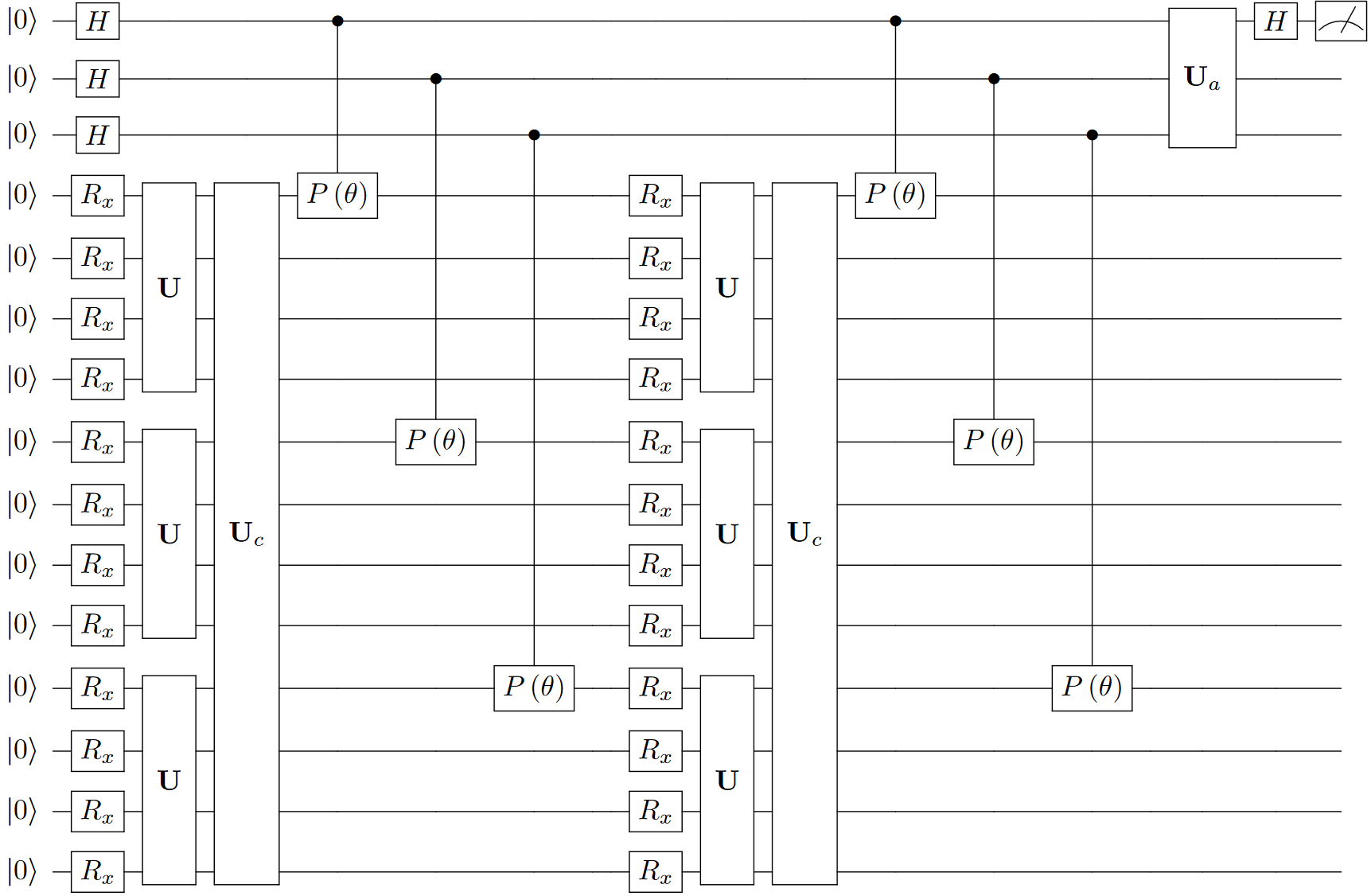}
  \caption{Parallel Channel Overwrite - Topologically Considerate (PCO-T) circuit with three registers using a 2 $\times$ 2 filter applied to six channels. $\bold{U}_a$ (Appendix Figure \ref{fig:inter_ancilla_a}) is a block of unitary gates to entangle the ancilla qubits.}
  \label{fig:PCO_T_circuit}
\end{figure}

The circuit construction put forth in Section \ref{DRPC_construction} is generalizable to $C$ channels and $R$ registers, however it becomes difficult to control each register from the same ancilla qubit as $R$ becomes large. To account for this difficulty, we present an additional method that mitigates the topological challenges of the hardware being used by introducing additional ancilla qubits. In this method, the desired number of controlled phase gates can be applied to each ancilla qubit. With this Parallel Channel Overwrite - Topologically Considerate (PCO-T-QCNN) method, the ancilla qubits are entangled with learnable unitaries at the conclusion of all controlled phase gates. The full circuit is pictured in Figure \ref{fig:PCO_T_circuit}. The total number of qubits in this method is $RF^2+A$, where $A$ is the number of ancilla qubits chosen to decrease the degree of required connectivity.

\subsection{Weighted Expectation Value Method}

The Weighted Expectation Value (WEV)- QCNN method adds a hybrid component to classically learn inter-channel features. The quantum convolutions are performed in the traditional sense of passing the filter over each channel individually. After the expectation values are acquired, we apply a classical weight and bias before summing each value to generate the corresponding output:

\begin{equation}\label{eq:25}
    \Tilde{x}_{i',j'} = \sum_{c=1}^{C}\bra{\Psi_{i,j,c}}\sigma_z\ket{\Psi_{i,j,c}} \cdot  w_{i,j,c} + b_{i,j,c}
\end{equation}
$\Tilde{x}_{i',j'}$ is a single classical output data point given by the quantum convolution. $\Psi_{i,j}$ is the wavefunction after the quantum convolution. This method is ideal for few qubits, few gates, and shallow circuit depth.

\subsection{Convolutional Unitary Blocks}

\begin{figure}[H]
  \centering
  \includegraphics[width=0.45\textwidth]{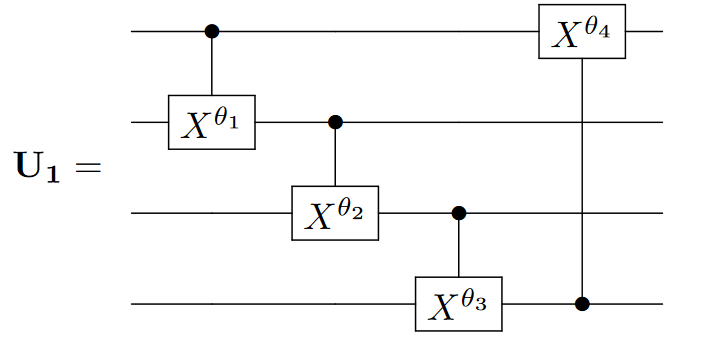}
  \includegraphics[width=0.45\textwidth]{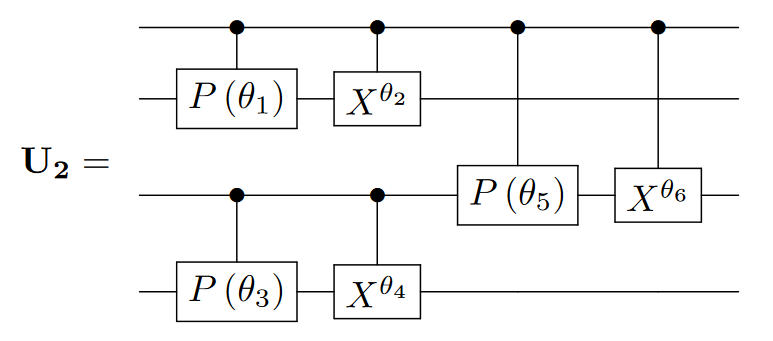}
  \caption{The $\bold{U_1}$ and $\bold{U_2}$ circuits used in this work, visualized for a 2 $\times$ 2 filter.}
  \label{fig:U_circuit}
\end{figure}

The unitary blocks used to perform the convolutions are designed to demonstrate the functionality of the proposed methods. Unitary blocks $\bold{U}_1$ and $\bold{U}_2$ are shown in Figure \ref{fig:U_circuit}, and are based on \cite{schuld2020circuit} and \cite{Oh2020}, respectively. The gate $X^\theta$ represents a power of an $X$ gate, and is described by Equation \ref{xpowgate}. The entangling unitary blocks $\bold{U}_c$ and $\bold{U}_a$ used in the PCO and PCO-T circuits are detailed in the Appendix.

\begin{equation}\label{xpowgate}
    X^\theta = 
    \begin{bmatrix}
        e^{\frac{i\pi\theta}{2}}\cos\left(\frac{\pi \theta}{2}\right) & -ie^{\frac{i\pi\theta}{2}}\sin\left(\frac{\pi \theta}{2}\right) \\
        -ie^{\frac{i\pi\theta}{2}}\sin\left(\frac{\pi \theta}{2}\right) & e^{\frac{i\pi\theta}{2}}\cos\left(\frac{\pi \theta}{2}\right)
    \end{bmatrix}
\end{equation}

\section{Methods}\label{sec5}
\subsection{Hardware}
All scripting was done in Python 3.9.13. The quantum simulation package Cirq \cite{cirq} version 0.13.1 was used, as well as Tensorflow 2.7.0 \cite{tensorflow2015-whitepaper} and Tensorflow Quantum 0.7.2 \cite{Broughton2020}. An Intel i7-13700KF CPU, 12GB Nvidia GeForce RTX 3080Ti GPU, and 64GB of 3600MHz CL18 RAM were used for all computations. 
\subsection{Datasets}
The CIFAR-10 dataset (Section \ref{cifar_10_dataset}) was loaded with Tensorflow Keras, and the pixel values were normalized between 0 and 1. The training and testing data were segregated by class, and the first 500 data points in each class of the training data were used as the training set. The first 100 data points in each class of the test data were used as the test set. Both sets were then shuffled using scikit-learn \cite{scikit-learn}. After building the training and test sets, all images were downsized from 32 $\times$ 32 $\times$ 3 to 10 $\times$ 10 $\times$ 3 using bilinear interpolation. 

The \textit{noisy color dataset} (Section \ref{noisy_color_dataset}) was created by making a colored 10 $\times$ 10 - pixel square in Microsoft Paint using the RGB values specified in Table \ref{noisy_colors_values} and then images were saved as PNG files. For each color, 400 replicas of the images were loaded with Tensorflow, and 20\% of the pixels were randomly corrupted in each by setting the pixel value to 0, 0, 0. The \textit{noisy colors with shapes dataset}  (Section \ref{noisy_color_shapes_dataset}) was prepared analogously, where the shapes were drawn by setting the pixel values to white (255, 255, 255). Each channel in both of these RGB synthetic datasets was normalized to $\frac{c}{C}$, where $c$ is the index of the channel and $C$ is the total number of channels. This allows the model to better distinguish between channels in the Bloch sphere.

The \textit{synthetic 12-channel dataset} (Section \ref{high_channel_dataset}) was created by initializing a 10 $\times$ 10 $\times$ 12 tensor, where all elements were populated with uniformly distributed random values between 0 and 1. 0.5 was added to every element in the three specified channels that comprised a class, where $i$ class is defined by adding 0.5 to channels $i$ to $i+2$. 100 replicas of each class were generated to serve as training data, and 20 replicas of each class were generated to serve as test data. 

\subsection{Machine Learning}

The learnable parameters in the quantum circuits were initialized using the Xavier method \cite{glorot2010}. For the CIFAR-10 and synthetic RGB datasets (all but the \textit{synthetic 12-channel dataset}), the classical weights and biases associated with the expectation values in the WEV-QCNN were initialized with a random normal distribution centered at 1.0 and 0.0, respectively, with both distributions having a standard deviation of 0.1. Xavier initialization was also used for the classical weights and biases when training on the \textit{synthetic 12-channel dataset}. Categorical cross-entropy was used as the loss function, and Adam optimizer \cite{Kingma2014} was used to update the weights. A constant learning rate of 0.001 and was used for all datasets, except for the \textit{synthetic 12-channel dataset} which used a learning rate of 0.01. The exponential decay rate for the first and second moment estimations were set to 0.9 and 0.999, respectively for all training. Epsilon was set to $1.0 \times 10^{-7}$. The output of each hidden layer in the model architecture was activated with ReLU, and Softmax was applied to the final output layer to obtain probabilities. The classical architecture was the same for all models. Accuracy was calculated as the ratio of images the model classified correctly to the number of total images the model attempted to classify.

\begin{table}
  \begin{minipage}{.5\linewidth}
    \centering
    
    \begin{tabular}{@{}cc@{}}
      \toprule
      Number of Classes & Class That is Added \\
      \midrule
      2 & frog, ship \\
      3 & 2 + automobile \\
      4 & 3 + truck \\
      5 & 4 + airplane \\
      6 & 5 + bird \\
      7 & 6 + cat \\
      8 & 7 + horse \\
      9 & 8 + dog \\
      10 & 9 + deer \\
      \bottomrule
    \end{tabular}
    \caption{Classes used in CIFAR-10}
    \label{cifar_classes_values}
  \end{minipage}%
  \hspace{1.5cm}
  \begin{minipage}{.5\linewidth}
  
    \centering
    \begin{tabular}{@{}cc@{}}
      \toprule
      Color & RGB Value \\
      \midrule
      Blue & 0, 0, 255 \\
      Green & 0, 255, 0 \\
      Red & 255, 0, 0 \\
      Cyan & 0, 255 ,255 \\
      Magenta & 255, 0, 255 \\
      Yellow & 255, 255, 0 \\
      Light Cyan & 128, 255, 255 \\
      Pink & 255, 128, 255 \\
      Light Yellow & 255, 255, 128 \\
      \bottomrule
    \end{tabular}
    \caption{\textit{Noisy Colors} RGB Values}
    \label{noisy_colors_values}
  \end{minipage}
\end{table}

\section{Results and Discussion}\label{sec3}

\subsection{Architecture}

\begin{figure}
  \centering
  \includegraphics[width=0.9\textwidth]{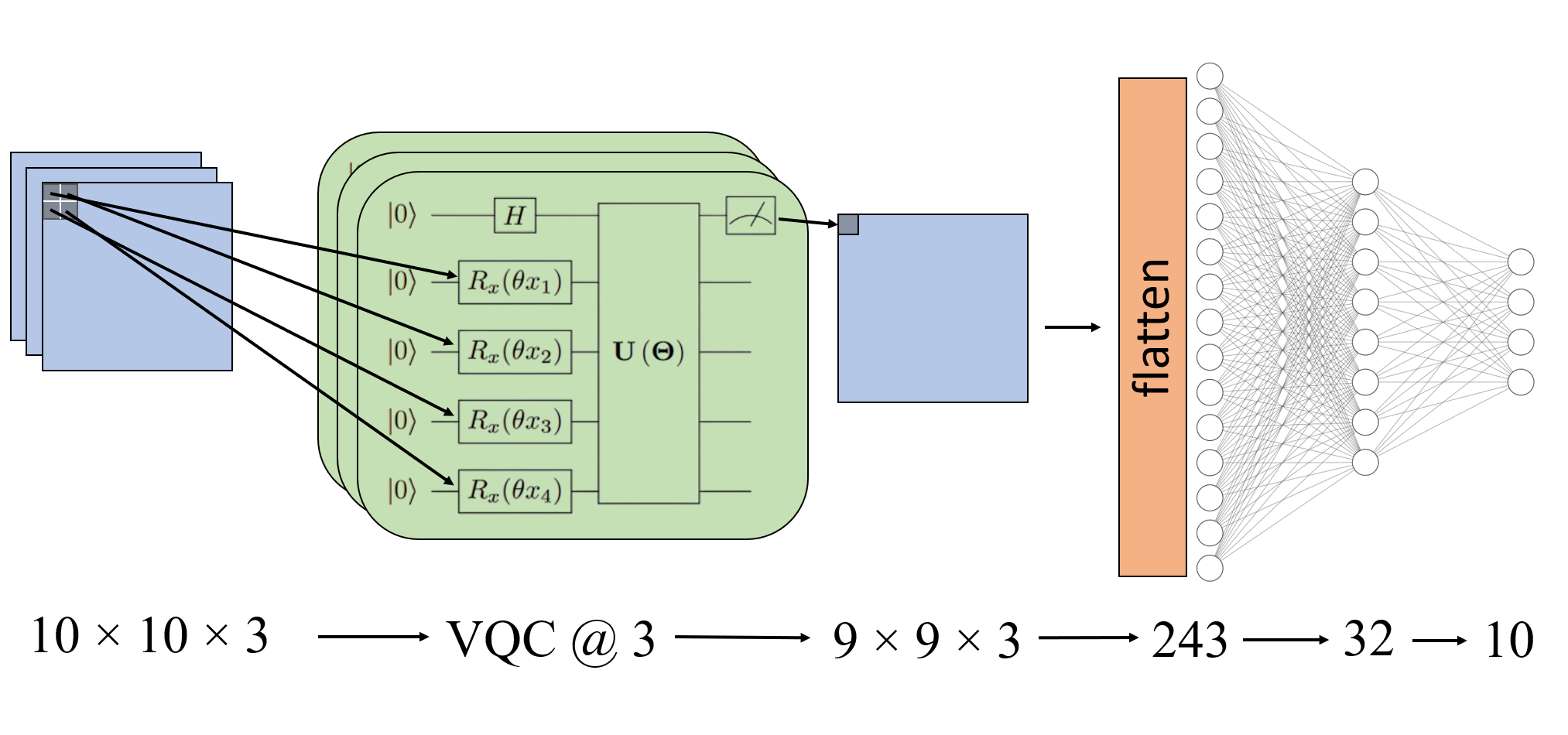}
  \caption{Hybrid quantum-classical machine learning architecture used to evaluate each of the proposed circuits.}
  \label{fig:architecture}
\end{figure}

We evaluate the proposed quantum circuits as components of a hybrid quantum-classical machine learning architecture for image classification. A quantum circuit is used to perform the convolution over the embedded input pixels of the image. This builds the output feature map, and a classical two-layer feed-forward network is used to obtain the probabilities for each possible class. This architecture is portrayed in Figure \ref{fig:architecture}.

\subsection{Datasets}

We utilize the CIFAR-10 dataset, as well as three sets of synthetic data designed to isolate the ability of our circuits to learn inter-channel information. All models are trained for 20 epochs and the hyperparameters used for each dataset are not changed between models.
\subsubsection{CIFAR-10}\label{cifar_10_dataset}

The Canadian Institute for Advanced Research, 10 classes (CIFAR-10) \cite{Krizhevsky2009} is a dataset of images that has become a prominent benchmark for image recognition models. This dataset consists of 50,000 training and 10,000 testing RGB images of ten different classes: airplane, automobile, bird, cat, dog, deer, frog, horse, ship, and truck. Each image is 32 $\times$ 32 pixels (in this work, we rescale them to 10 $\times$ 10 pixels). Unless otherwise specified, we train on 500 randomly selected images from each class, and test on 100 randomly selected images from each class that do not appear in the training set.

\begin{figure}[H]
\centering
    \begin{subfigure}{.33\textwidth}
      \centering
      \includegraphics[width=0.4\textwidth]{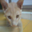}
      \caption{Cat}
      \label{fig:cat}
    \end{subfigure}%
    \begin{subfigure}{.33\textwidth}
      \centering
      \includegraphics[width=0.4\textwidth]{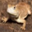}
      \caption{Frog}
      \label{fig:frog}
    \end{subfigure}%
    \begin{subfigure}{.33\textwidth}
      \centering
      \includegraphics[width=0.4\textwidth]{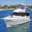}
      \caption{Ship}
      \label{fig:ship}
    \end{subfigure}

\caption{Three representative full-resolution (32 $\times$ 32 pixel) images from the CIFAR-10 dataset.}
\label{fig:cifar10}
\end{figure}

\subsubsection{Synthetic Noisy Colors}\label{noisy_color_dataset}

\begin{figure}[H]
\centering
    \begin{subfigure}{.33\textwidth}
      \centering
      \includegraphics[width=0.5\textwidth]{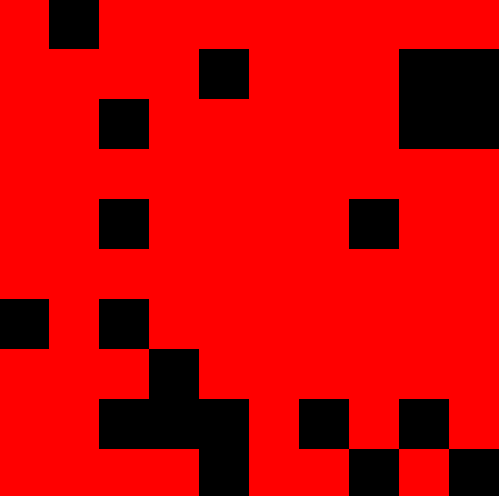}
      \caption{Red}
      \label{fig:Noisy_Red}
    \end{subfigure}%
    \begin{subfigure}{.33\textwidth}
      \centering
      \includegraphics[width=0.5\textwidth]{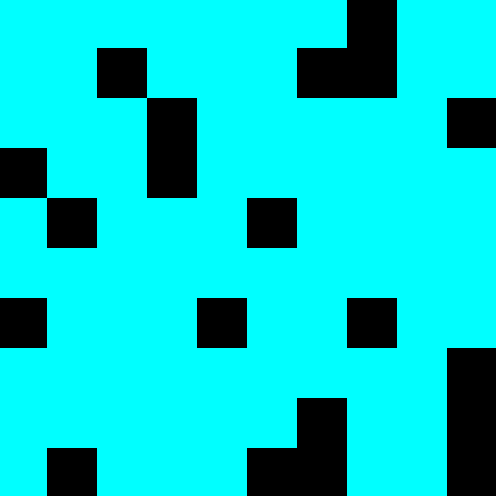}
      \caption{Cyan}
      \label{fig:Noisy_Cyan}
    \end{subfigure}%
    \begin{subfigure}{.33\textwidth}
      \centering
      \includegraphics[width=0.5\textwidth]{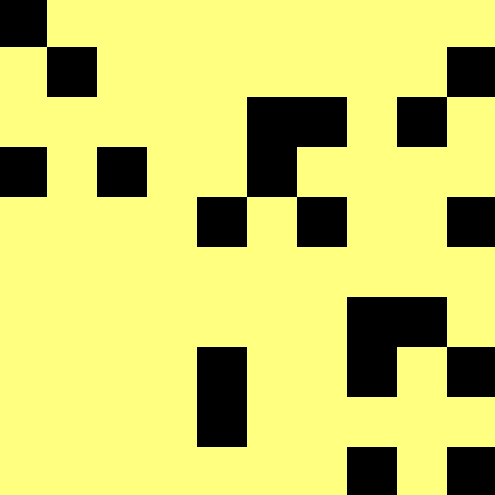}
      \caption{Light Yellow}
      \label{fig:Noisy_Light_Yellow}
    \end{subfigure}

\caption{Three representative noisy color data points from the data set. Each color in the figure has a different number of populated channels: red (255, 0, 0), cyan (0, 255, 255), and light yellow (255, 255, 128).}
\label{fig:Noisy_Colors}
\end{figure}

10 $\times$ 10 pixel images of nine colors are created (blue, green, red, cyan, magenta, yellow, light cyan, pink, and light yellow), and  20\% of these pixels are randomly selected and corrupted by setting all  channel values of the corrupted pixel to 0. Three representative images are shown in Figure \ref{fig:Noisy_Colors}. In total, there are 400 images for each color (3,600 total images), where 80\% of these images were used for training and 20\% are used for testing.

\subsubsection{Synthetic Noisy Patterned Colors}\label{noisy_color_shapes_dataset}
\begin{figure}[H]
\centering
    \begin{subfigure}{.25\textwidth}
      \centering
      \includegraphics[width=0.7\textwidth]{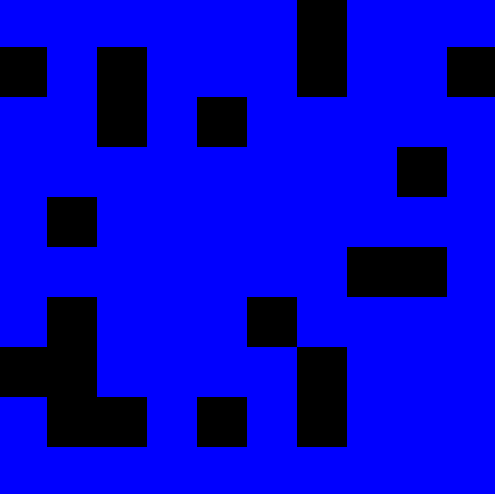}
      \caption{Blue, No design}
      \label{fig:Noisy_Blue}
    \end{subfigure}%
    \begin{subfigure}{.25\textwidth}
      \centering
      \includegraphics[width=0.7\textwidth]{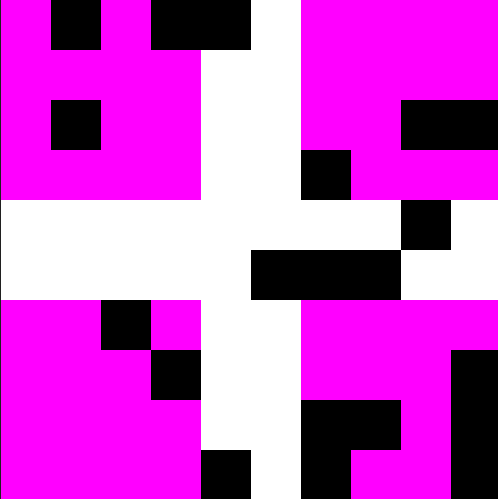}
      \caption{Magenta, Cross}
      \label{fig:Noisy_Cross_Magenta}
    \end{subfigure}%
    \begin{subfigure}{.25\textwidth}
      \centering
      \includegraphics[width=0.7\textwidth]{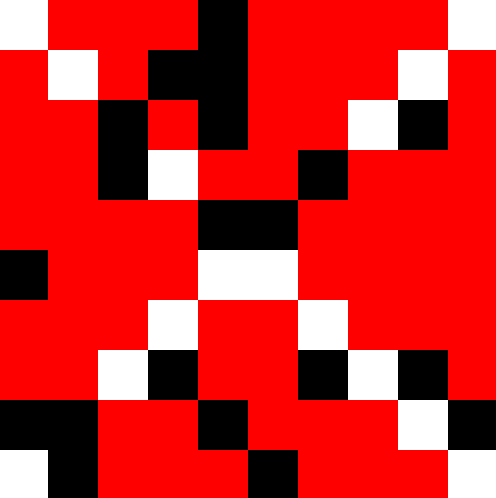}
      \caption{Red, X}
      \label{fig:Noisy_Red_X}
    \end{subfigure}%
    \begin{subfigure}{.25\textwidth}
      \centering
      \includegraphics[width=0.7\textwidth]{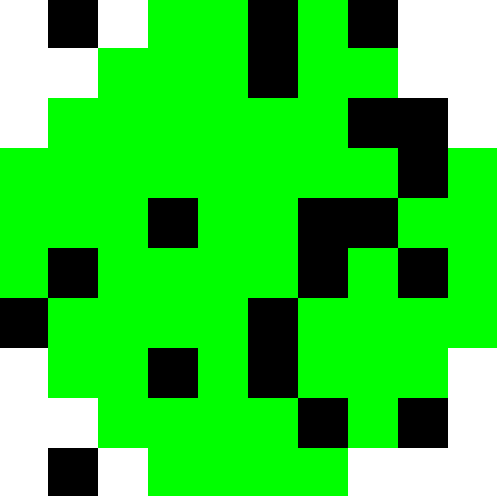}
      \caption{Green, Rounded}
      \label{fig:Noisy_Rounded_Green}
    \end{subfigure}

\caption{Each design shape of patterned noisy color from the dataset. The designs are created using white (255, 255, 255), and then subsequently subjected to the random corruption.}
\label{fig:Noisy_Colors_Shapes}
\end{figure}

To demonstrate that this method also captures intra-channel features, four different design shapes are drawn on the synthetic colors in white before corruption. In this dataset six base colors (blue, green, red, cyan, magenta, and yellow) are used, for a total of 24 classes, four of which are shown in Figure \ref{fig:Noisy_Colors_Shapes}. We create 400 10 $\times$ 10 training images of each class, and use 80\% for training and 20\% for testing.

\subsubsection{Synthetic High-Channel Data}\label{high_channel_dataset}
\begin{figure}[H]
\centering
    \begin{subfigure}{.33\textwidth}
      \centering
      \includegraphics[width=0.85\textwidth]{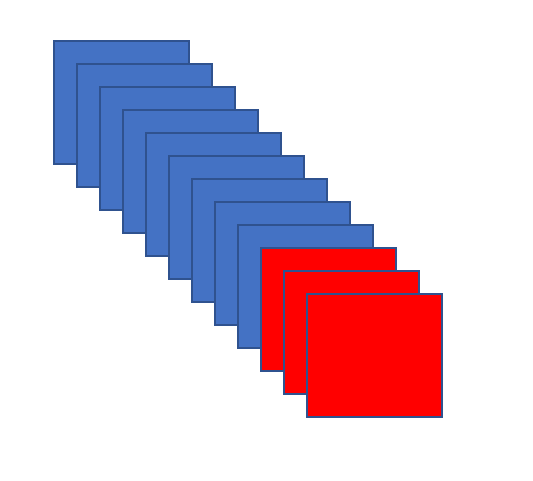}
      \caption{First class}
      \label{fig:channels_1}
    \end{subfigure}%
    \begin{subfigure}{.33\textwidth}
      \centering
      \includegraphics[width=0.70\textwidth]{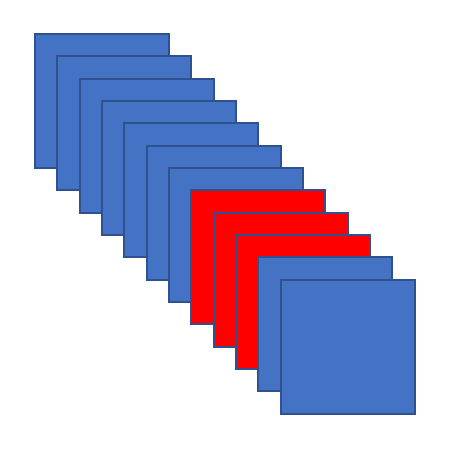}
      \caption{Third class}
      \label{fig:channels_3}
    \end{subfigure}%
    \begin{subfigure}{.33\textwidth}
      \centering
      \includegraphics[width=0.85\textwidth]{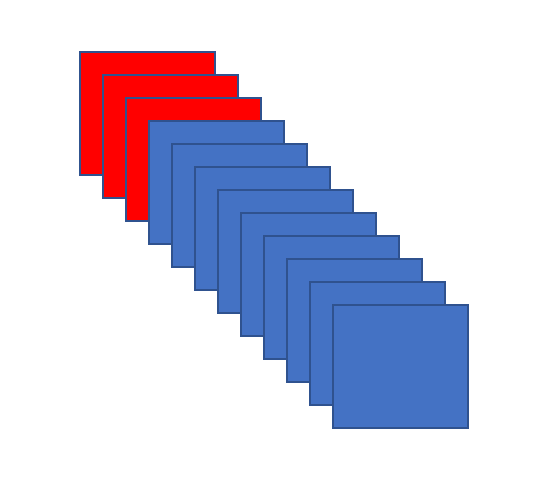}
      \caption{Tenth class}
      \label{fig:channels_10}
    \end{subfigure}

\caption{Synthetic high-channel data. Each element in the blue colored channels are random numbers between 0 and 1. Each element in the red colored channels are random numbers between 0.5 and 1.5.}
\label{fig:channels}
\end{figure}

We create a 12-channel dataset consisting of tensors of shape 10 $\times$ 10 $\times$ 12 that are each populated with a uniform random distribution of numbers $\in \lbrack 0,1)$. Each of the 10 classes is defined by the addition of 0.5 to three distinct channels of the data, and is shown in Figure \ref{fig:channels}.

\subsection{CIFAR-10 Dataset Results}

\begin{figure}[!hbt]
  \centering
  \captionsetup[subfigure]{position=top} 
  
  \makebox[\textwidth][c]{
    \begin{subfigure}{0.45\textwidth}
      \centering
      
      \includegraphics[width=\linewidth]{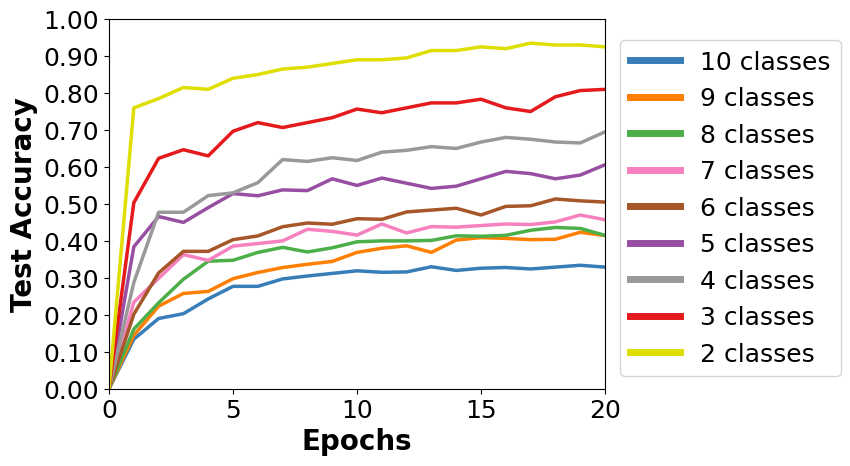}
      \caption{CO-QCNN}
      \label{fig:subfig1}
    \end{subfigure}
    \hspace{0.05\textwidth}
    \begin{subfigure}{0.45\textwidth}
      \centering
      
      \includegraphics[width=\linewidth]{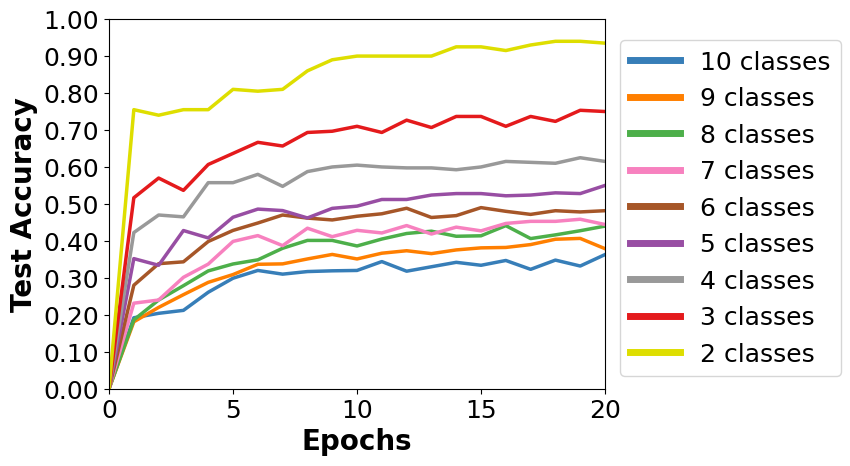}
      \caption{PCO-QCNN}
      \label{fig:subfig2}
    \end{subfigure}
  }

  \vspace{0.05\textwidth}

  \makebox[\textwidth][c]{
    \begin{subfigure}{0.45\textwidth}
      \centering
      
      \includegraphics[width=\linewidth]{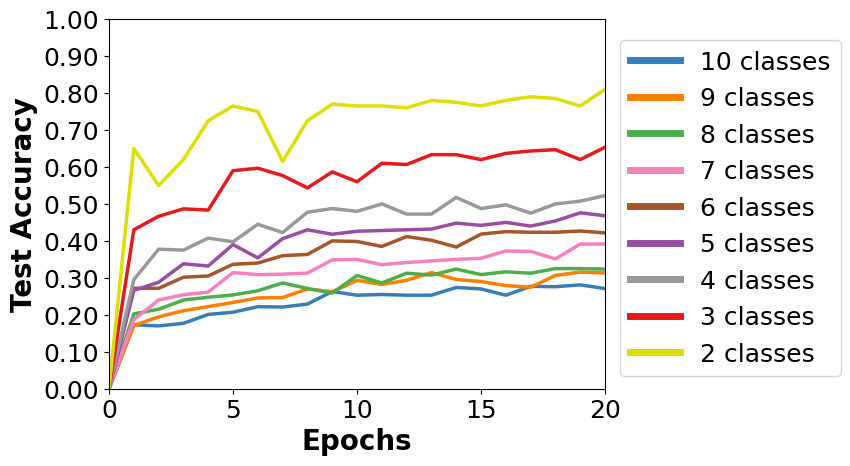}
      \caption{PCO-T-QCNN}
      \label{fig:subfig3}
    \end{subfigure}
    \hspace{0.05\textwidth}
    \begin{subfigure}{0.45\textwidth}
      \centering
      
      \includegraphics[width=\linewidth]{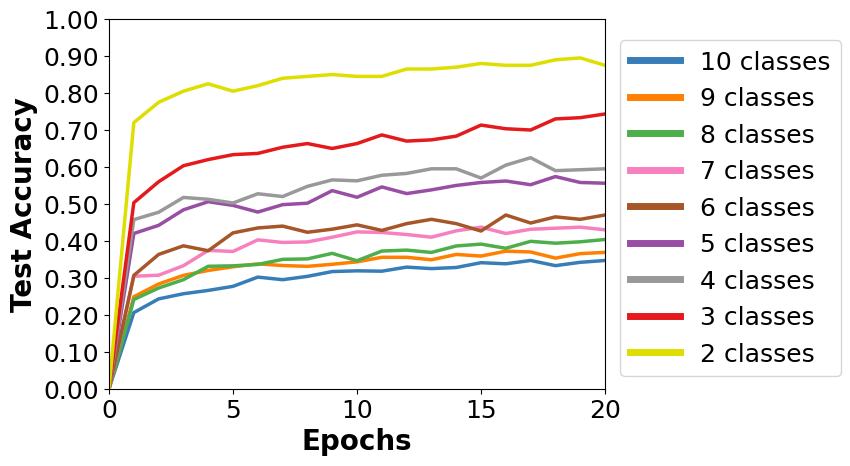}
      \caption{WEV-QCNN}
      \label{fig:subfig4}
    \end{subfigure}
  }

  \vspace{0.05\textwidth}

  \makebox[\textwidth][c]{
    \begin{subfigure}{0.45\textwidth}
      \centering
      
      \includegraphics[width=\linewidth]{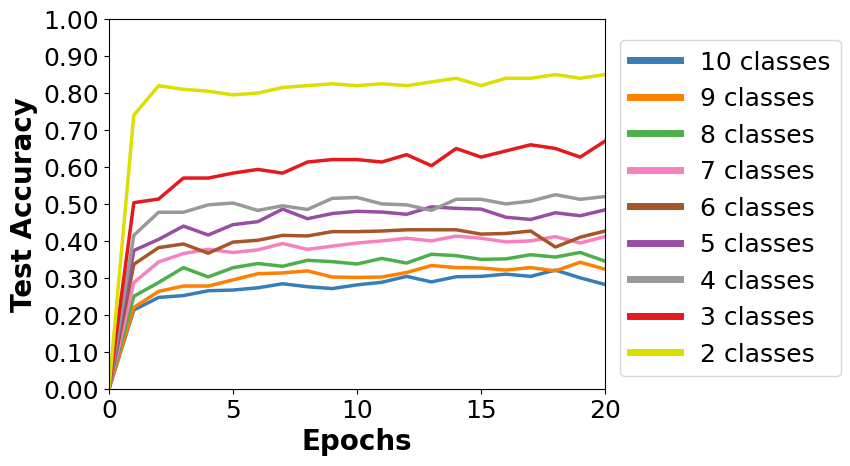}
      \caption{Control}
      \label{fig:subfig5}
    \end{subfigure}
  }

  \caption{Test set accuracy as a function of training epochs for the CIFAR dataset. Channel Overwrite (CO) quantum convolutional neural network (QCNN), Parallel CO (PCO)-QCNN, PCO - topologically considerate (PCO-T)-QCNN, weighted expectation value (WEV)-QCNN, and a control QCNN are evaluated for two to ten classes on the CIFAR dataset. Quantum convolutions are performed with $\bold{U}_1$.}
  \label{fig:CIFAR_classes}
\end{figure}

All four proposed methods CO-QCNN, PCO-QCNN, PCO-T-QCNN, and WEV-QCNN are tested against a control QCNN. This control QCNN operates by passing the quantum filter over each channel individually, acquiring an expectation value, and summing the expectation values to produce the final output. 

\begin{figure}
\centering
    \begin{subfigure}{.50\textwidth}
      \centering
      \includegraphics[width=0.9\textwidth]{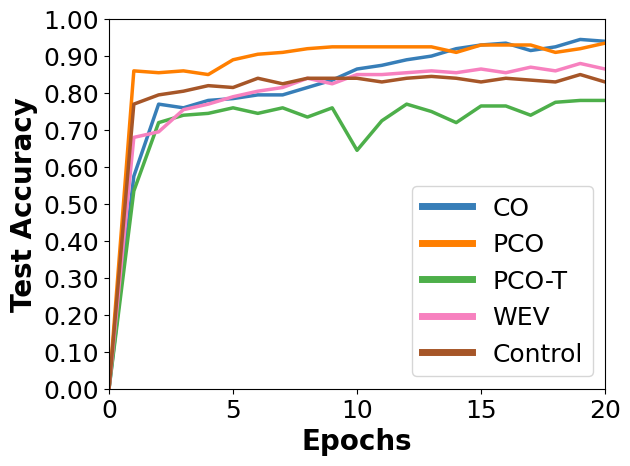}
      \caption{CIFAR-2 Classification Accuracy}
      \label{fig:cifar_2}
    \end{subfigure}%
    \begin{subfigure}{.50\textwidth}
      \centering
      \includegraphics[width=0.9\textwidth]{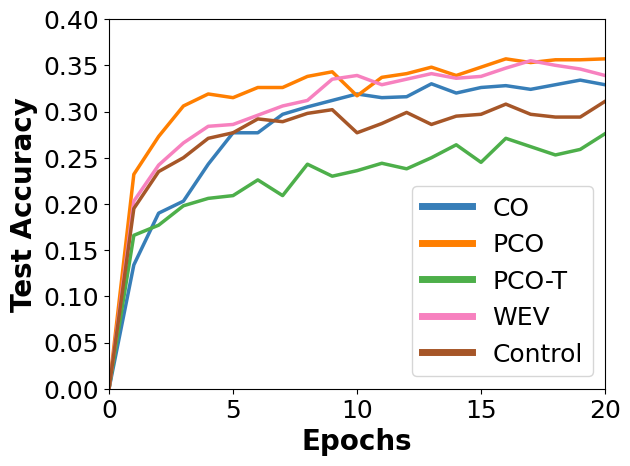}
      \caption{CIFAR-10 Classification Accuracy}
      \label{fig:cifar_10}
    \end{subfigure}

\caption{Test set accuracy as a function of training epochs for the Channel Overwrite (CO) quantum convolutional neural network (QCNN), Parallel CO (PCO)-QCNN, PCO topologically considerate (PCO-T)-QCNN, weighted expectation value (WEV)-QCNN, and a control QCNN for binary and ten-member classification. Quantum convolutions shown are performed with $\bold{U}_2$.}
\label{fig:CIFAR_circuits}
\end{figure}

To comprehensively evaluate the proposed QCNNs on the CIFAR-10 dataset, each model is evaluated on an \textit{n}-member classification task, where \textit{n} (two - ten) classes are contained in the training and test sets. We begin with ten classes and iteratively remove the remaining class that the CO-QCNN classifies with the least accuracy, performing a classification at each iteration. The PCO-QCNN using $\bold{U}_2$ achieves the highest binary accuracy of 94.5\%, compared to 85.0\% with the control QCNN. When trained on all ten classes, the CO-QCNN was able to achieve 34.9\% while the control QCNN produced 28.2\% accuracy using $\bold{U}_1$. The performance of each method on variable numbers of classes is shown in Figure \ref{fig:CIFAR_classes}. A comparison between each circuit is displayed for each type of classification in Figure \ref{fig:CIFAR_circuits}. For all classifications, the CO-QCNN, PCO-QCNN, and WEV-QCNN outperformed the control QCNN and achieve state-of-the-art accuracy compared to current quantum neural networks (Table \ref{full_cifar}).

\begin{table}[hb]
\caption{Full 60,000 Images CIFAR-10 Image Recognition}\label{full_cifar}%
\begin{tabular}{@{}cc@{}}
\toprule
Quantum Model & Test Accuracy\\
\midrule
Quanvolutional Neural Network\footnotemark[1] \cite{Mari_2021,Riaz2023} & 34.9\% \\
Neural Network with Quantum Entanglement\footnotemark[1] \cite{Riaz2023} & 36.0\% \\
Flat Quantum Convolutional Ansatz\footnotemark[2] \cite{Jing2022} & 41.8\% \\
CO-QCNN (U1/U2) & 43.1/45.1\% \\
PCO-QCNN (U1/U2) & 44.9/44.4\%\\
WEV-QCNN (U1/U2) & 44.0/44.3\%\\
\midrule
Classical Model & Test Accuracy\\
\midrule
CNN\footnotemark[3] & 64.6\% \\
CNN-P\footnotemark[3] & 51.2\% \\
\botrule
\end{tabular}
\footnotetext[1]{Models make no attempt to learn inter-channel information, and instead converts the images to grayscale.}
\footnotetext[2]{Model is tested with images from the training data. This accuracy is a reproduction trained and tested on the full CIFAR-10 data using their exact architecture and circuit.}
\footnotetext[3]{Both architectures are shown in Figures \ref{fig:CNN_arch} and \ref{fig:CNN-P_arch} in the Appendix. CNN-P is a CNN with the same number of parameters as our work.}
\end{table}

\subsection{Synthetic Dataset Results}
In an attempt to demonstrate that the increased accuracy of the CO-QCNN, PCO-QCNN, and WEV-QCNN models can be attributed to their ability to extract features across the channel dimension, all proposed methods and the control QCNN are evaluated with synthetic datasets that emphasize inter-channel relationships. These synthetic datasets place all patterns to be learned in the channel dimension only, rather than both the spatial and channel dimension of the data. In an RGB image, the color is explicitly defined by the value of each channel. In the synthetic 12-channel dataset, the classes are determined by which channels contains higher values than the others. By relegating all important learnable patterns to the channel dimension, the proposed models' abilities to extract inter-channel features are demonstrated. 
 
All proposed methods except PCO-T-QCNN are able to achieve 98\% accuracy, with CO-QCNN achieving 100\% accuracy on the \textit{synthetic 12-channel dataset}. When predicting the color of noisy RGB images, the proposed models demonstrate the ability to learn, with the CO-QCNN and PCO-QCNN achieving 100\% accuracy. The control QCNN is shown to perform much worse. The results for the \textit{noisy colors with shapes dataset} are similar to those of the \textit{noisy color dataset}.

\begin{figure}
\centering
    \begin{subfigure}{.50\textwidth}
      \centering
      \includegraphics[width=0.9\textwidth]{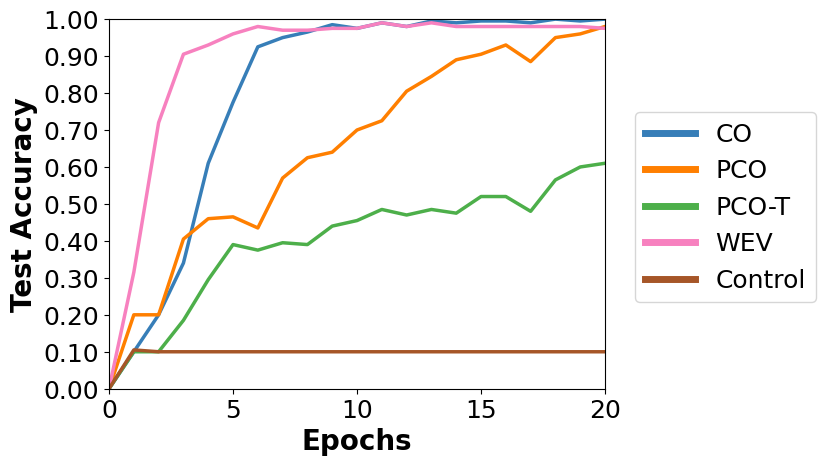}
      \caption{12-Channel Dataset Accuracy}
      \label{fig:high_channel_acc}
    \end{subfigure}%
    \begin{subfigure}{.50\textwidth}
      \centering
      \includegraphics[width=0.9\textwidth]{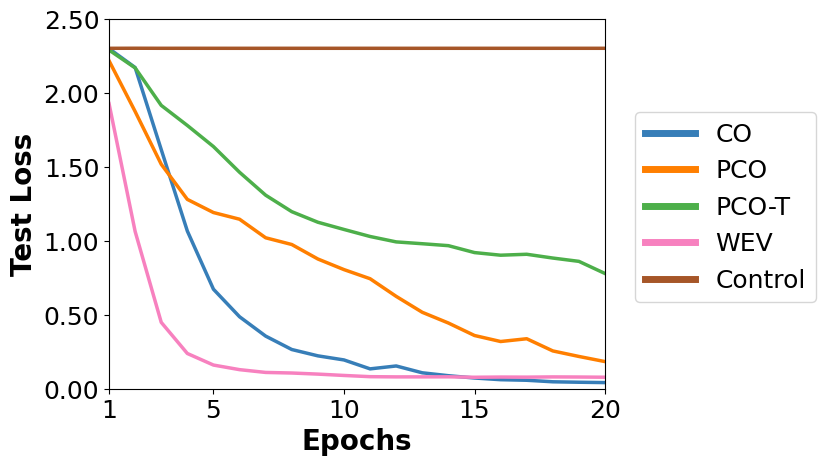}
      \caption{12-Channel Dataset Loss}
      \label{fig:high_channel_loss}
    \end{subfigure}

\caption{Test set accuracy as a function of training epochs for the Channel Overwrite (CO) quantum convolutional neural network (QCNN), Parallel CO (PCO)-QCNN, PCO topologically considerate (PCO-T)-QCNN, weighted expectation value (WEV)-QCNN, and a control QCNN for the classification of \textit{synthetic 12-channel data}. Quantum convolutions shown are performed with $\bold{U}_2$.}
\end{figure}

\begin{figure}[!hbt]
  \centering
  \captionsetup[subfigure]{position=top} 

  \makebox[\textwidth][c]{
    \begin{subfigure}{0.45\textwidth}
      \centering
      \includegraphics[width=\linewidth]{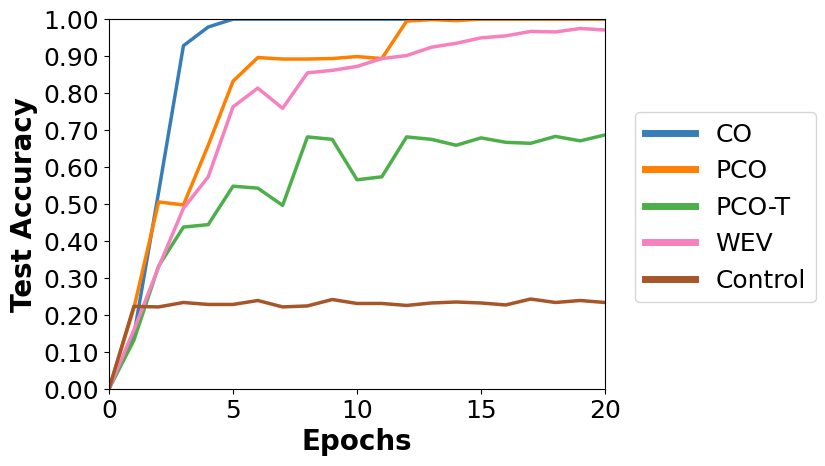}
      \caption{Noisy Colors Accuracy}
      \label{fig:colors_acc}
    \end{subfigure}
    \hspace{0.05\textwidth}
    \begin{subfigure}{0.45\textwidth}
      \centering
      \includegraphics[width=\linewidth]{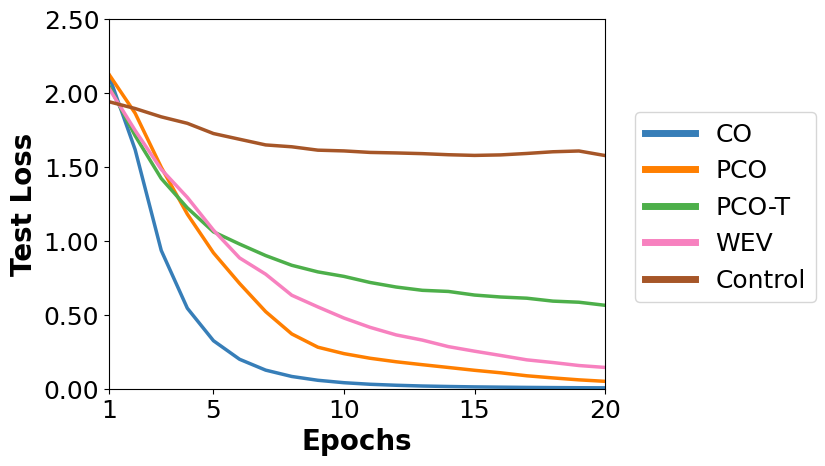}
      \caption{Noisy Colors Loss}
      \label{fig:colors_loss}
    \end{subfigure}
  }

  \vspace{0.05\textwidth}

  \makebox[\textwidth][c]{
    \begin{subfigure}{0.45\textwidth}
      \centering
      \includegraphics[width=\linewidth]{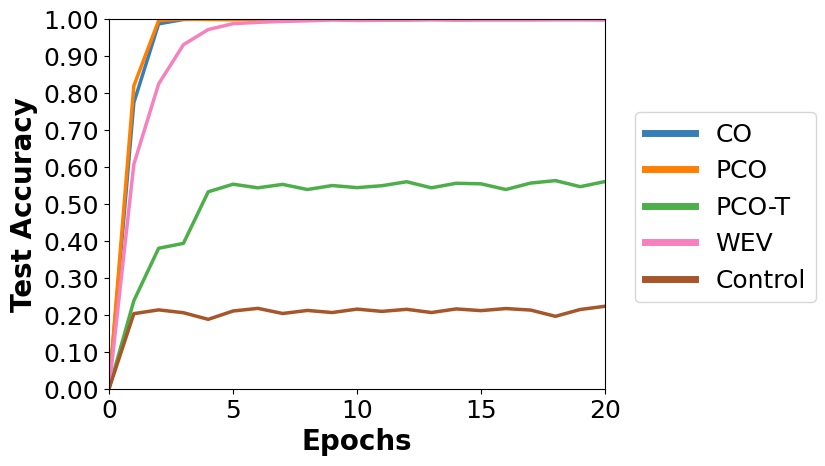}
      \caption{Noisy Colors with Shapes Loss}
      \label{fig:shapes_acc}
    \end{subfigure}
    \hspace{0.05\textwidth}
    \begin{subfigure}{0.45\textwidth}
      \centering
      \includegraphics[width=\linewidth]{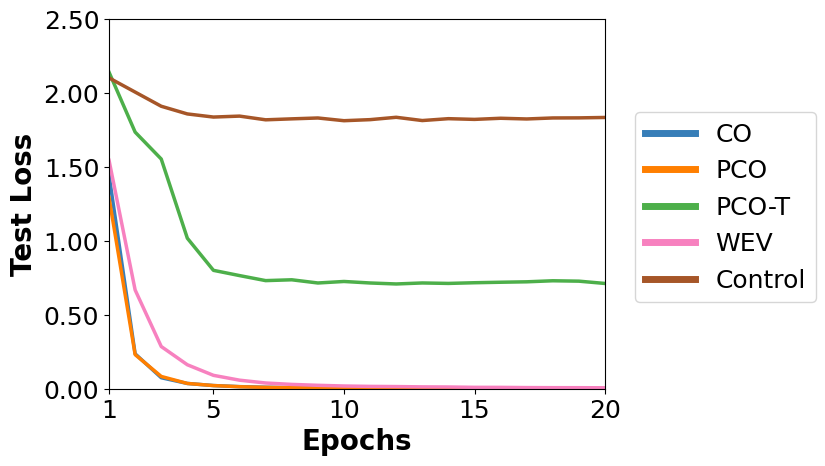}
      \caption{Noisy Colors with Shapes Loss}
      \label{fig:shapes_loss}
    \end{subfigure}
  }

  \caption{Test set accuracy as a function of training epochs for the Channel Overwrite (CO) quantum convolutional neural network (QCNN), Parallel CO (PCO)-QCNN, PCO topologically considerate (PCO-T)-QCNN, weighted expectation value (WEV)-QCNN, and a control QCNN for the classification of synthetic RGB data. Quantum convolutions shown are performed with $\bold{U}_1$.}
  \label{fig:synthetic_results}
\end{figure}

\section{Conclusions}\label{sec4}
\begin{table}[h]
\caption{Proposed Quantum Circuit Complexities}\label{circuit_complexities}%
\begin{tabular}{@{}cccc@{}}
\toprule
Quantum Model & Dataset & Circuit Depth & Circuit Width\\
\midrule
\multirow{1}{8em}{FQConv\cite{Jing2022}} & CIFAR & 25 & 12\\
\midrule
\multirow{2}{8em}{CO-QCNN} & RGB Images & 19 & 5\\
& High Channel & 73 & 5\\
\midrule
\addlinespace[2ex]
\multirow{2}{8em}{PCO-QCNN} & RGB Images & 12 & 13\\
& High Channel & 42 & 13\\
\midrule
\addlinespace[2ex]
\multirow{2}{8em}{PCO-T-QCNN} & RGB Images & 13 & 15\\
& High Channel & 40 & 15\\
\midrule
\addlinespace[2ex]
\multirow{2}{8em}{WEV-QCNN} & RGB Images & 5& 4\\
& High Channel & 5& 4\\
\addlinespace[2ex]

\botrule
\end{tabular}

\end{table}
The proposed quantum circuits allow our QCNNs to effectively learn inter-channel information, as shown through evaluation of the models on synthetic data that holds important patterns in the channel dimension. Moreover, the CO-QCNN, PCO-QCNN, and WEV-QCNN methods achieve state-of-the-art performance on the CIFAR-10 dataset compared to current quantum neural networks. In particular, the CO-QCNN and WEV-QCNN methods achieve a greater accuracy while  being computationally cheaper in both circuit depth and circuit width compared to this work's predecessor \cite{Jing2022}.

The poorer performance of the proposed PCO-T-QCNN suggests better efforts must be made to effectively correlate the multiple ancilla qubits. It is our hope that soon this method will lose relevance, as fully connected quantum devices are beginning to emerge, most recently Honeywell's 32-qubit trapped-ion quantum processor \cite{Moses2023}. As these models possess a heightened ability to learn inter-channel patterns, we wish to apply these to deeper layers in quantum-hybrid neural networks. For the CO-QCNN, PCO-QCNN, and PCO-T-QCNN, a controlled phase gate with only the first qubit of each working register and the ancilla qubit(s) is used . It is theorized higher accuracy may be achieved by performing controlled phase gates targeting all qubits in the working register. However as this work aims to create hardware considerate circuits, this is left to future projects. Since all expectation values are weighted in the WEV-QCNN method, it is proposed that as this method is parallelized across a Quantum Processing Unit the classical weights will learn which qubits are better performing than others and weight them less in the final output. This could serve as a quantum error corrective technique inherent to the WEV-QCNN method. Running this method on real quantum hardware would test this theory. Beyond image recognition, we envision that these methods can be applied to problems in the natural and physical sciences.

\backmatter

\section{Data Availability}
All of our software including that to generate the synthetic datasets and reproduce the results from this work are available as open source at \href{https://github.com/anthonysmaldone/QCNN-Multi-Channel-Supervised-Learning}{https://github.com/anthonysmaldone/QCNN-Multi-Channel-Supervised-Learning}.

\section{Declarations}
\subsection{Competing Interests}
The authors declare no competing financial interest.
\subsection{Author Contributions}
AMS, GWK designed research; AMS, GWK, developed software; AMS, GWK performed research; AMS, GWK, VSB analyzed data; and AMS, GWK wrote the paper. All authors have given approval to the final version of the manuscript.

\section{Acknowledgements}
We acknowledge the support provided by the NSF CCI grant (Award Number 2124511). 
\begin{appendices}
\section*{Appendix}
  \addcontentsline{toc}{section}{Appendix}

\begin{figure}[H]
  \centering
  \includegraphics[width=1.0\textwidth]{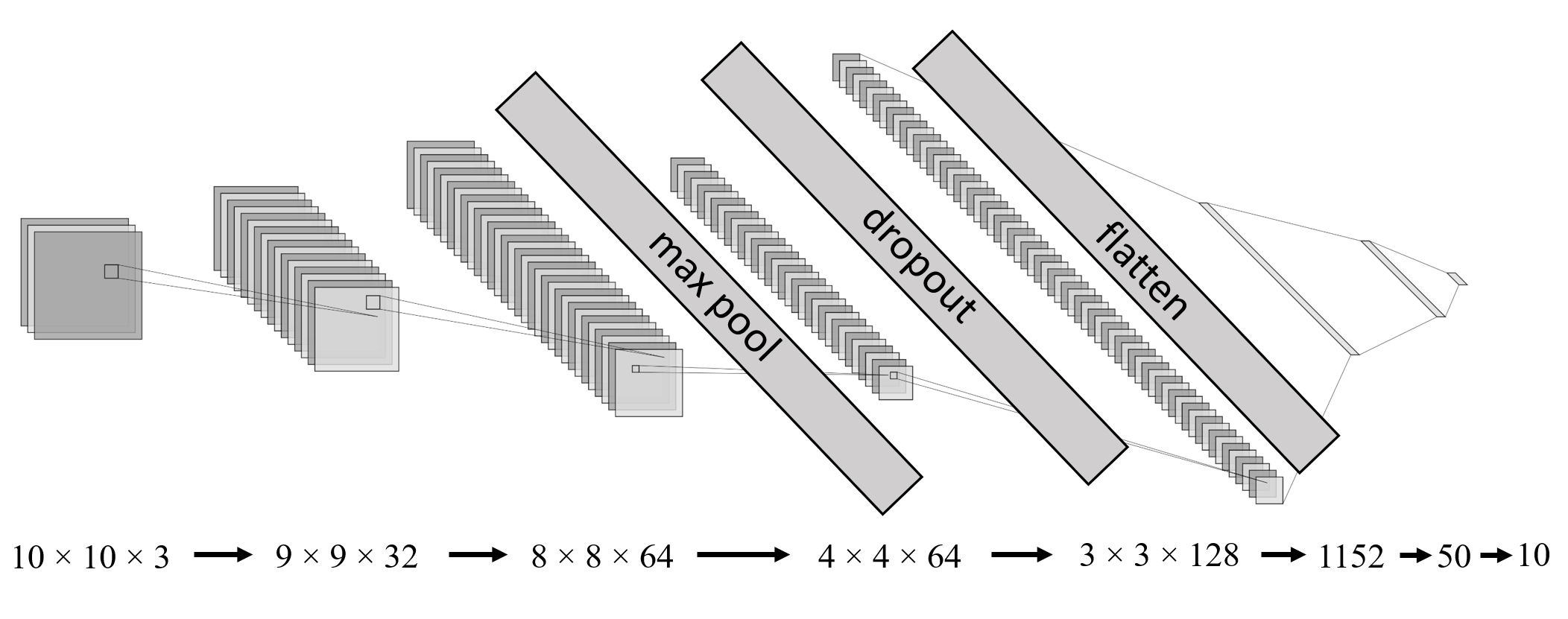}
  \caption{CNN Architecture}
  \label{fig:CNN_arch}
\end{figure}

\begin{figure}[H]
  \centering
  \includegraphics[width=1.0\textwidth]{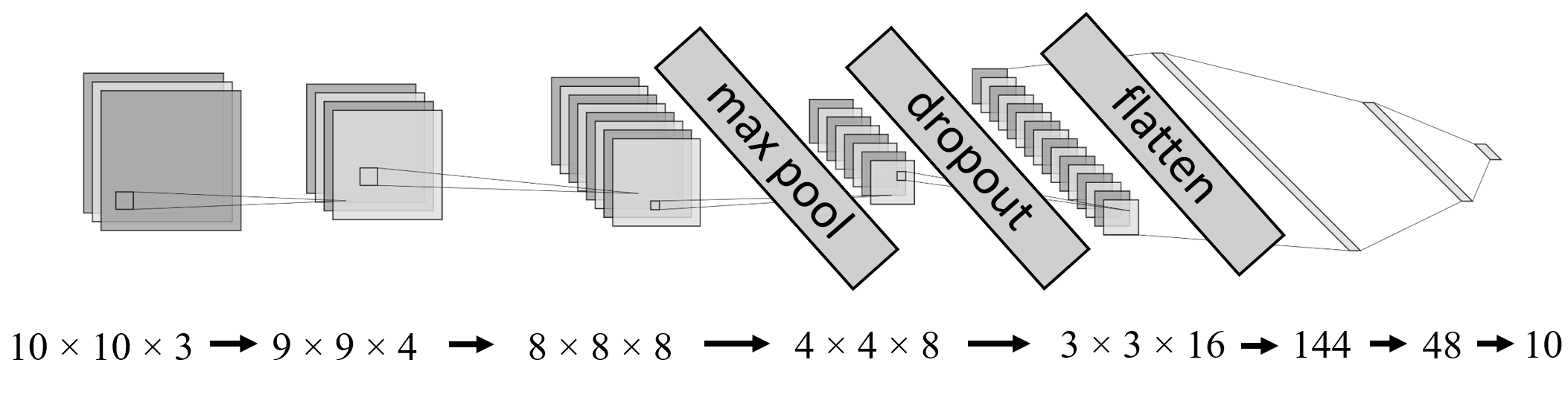}
  \caption{CNN-P Architecture}
  \label{fig:CNN-P_arch}
\end{figure}

\begin{figure}[H]
  \centering
  \includegraphics[width=0.4\textwidth]{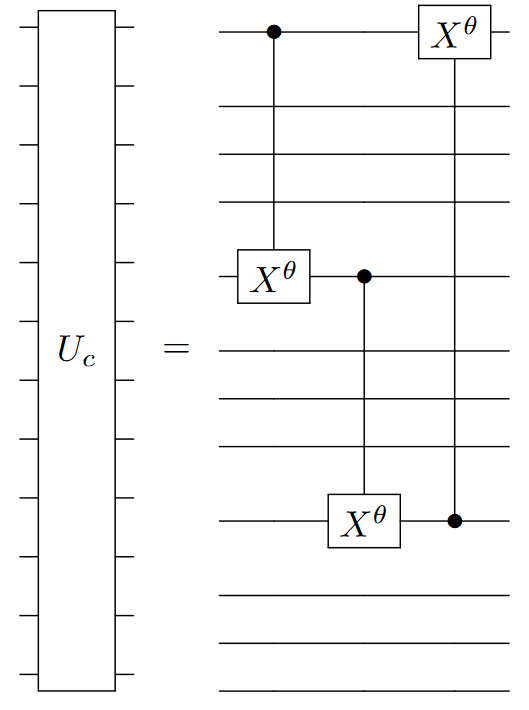}
  \caption{Unitary block used to entangle the working registers in the PCO and PCO-T methods.}
  \label{fig:inter_channel_u}
\end{figure}

\begin{figure}[H]
  \centering
  \includegraphics[width=0.5\textwidth]{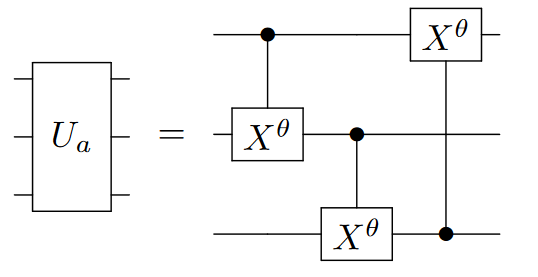}
  \caption{Unitary block used to entangle the ancilla register in the PCO-T method.}
  \label{fig:inter_ancilla_a}
\end{figure}




\end{appendices}


\bibliography{sn-bibliography}

\end{document}